\documentclass[journal]{IEEEtran}

\usepackage{epsfig,amsmath,amssymb,epsf,amsthm,scalefnt,multirow,subfig}
\usepackage{xcolor}
\usepackage{float}
\usepackage{cite}
\usepackage{psfrag}
\usepackage{graphics,amsmath,amssymb,graphicx,amsthm,scalefnt,multirow,dsfont,subfig}
\usepackage[linesnumbered,ruled,vlined]{algorithm2e}

\newtheorem*{lemma*}{Lemma}

\newtheorem{proposition}{Proposition}

  \def\cC{{\mathcal{C}}} 
   
\def\cI{{\mathcal{I}}}   
 \def\cN{{\mathcal{N}}} \def\cO{{\mathcal{O}}} 
   
  \def\cW{{\mathcal{W}}}

\def\bPhi{{\pmb{\Phi}}} \def\bphi{{\pmb{\phi}}}

\def\bTheta{{\pmb{\Theta}}}

\def\b0{{\pmb{0}}} 

   \def\bh{{\mathbf{h}}}
   
 \def\bn{{\mathbf{n}}}  
   
\def\bu{{\mathbf{u}}}   \def\bx{{\mathbf{x}}}
\def\by{{\mathbf{y}}}

   \def\bH{{\mathbf{H}}}
\def\bI{{\mathbf{I}}}  \def\bK{{\mathbf{K}}} 
 \def\bN{{\mathbf{N}}}  
 \def\bR{{\mathbf{R}}} \def\bS{{\mathbf{S}}} 
\def\bU{{\mathbf{U}}}   \def\bX{{\mathbf{X}}}
 \def\bZ{{\mathbf{Z}}}

\begin{document}	
	
\bstctlcite{IEEEexample:BSTcontrol}
\title{Channel Estimation for Reconfigurable Intelligent Surface with a few Active Elements}

\author{\IEEEauthorblockN{Gyoseung Lee, Hyeongtaek Lee, Jaeky Oh, Jaehoon Chung, and Junil Choi}
\thanks{This work was supported by LG Electronics Inc.; in part by the Ministry of Science and ICT (MSIT), South Korea, under the Information Technology Research Center (ITRC) Support Program supervised by the Institute of Information and Communications Technology Planning and Evaluation (IITP) under Grant IITP-2020-0-01787; and in part by the Korea Institute for Advancement of Technology (KIAT) grant funded by the Ministry of Trade, Industry and Energy (MOTIE) (P0022557).}
\thanks{Gyoseung Lee, Hyeongtaek Lee, and Junil Choi are with the School of Electrical Engineering, Korea Advanced Institute of Science and Technology (e-mail: \{iee4432; htlee8459; junil\}@kaist.ac.kr). Gyoseung Lee and Hyeongtaek Lee contributed equally to this work.}
\thanks{Jaeky Oh and Jaehoon Chung are with C\&M Standard Lab, ICT Technology Center, LG Electronics Inc. (e-mail: \{jaeky.oh; jaehoon.chung\}@lge.com).}}

\maketitle

\begin{abstract}
In this paper, a channel estimation technique for reconfigurable intelligent surface (RIS)-aided multi-user multiple-input single-output communication systems is proposed.
By deploying a small number of active elements at the RIS, the RIS can receive and process the training signals. Through the partial channel state information (CSI) obtained from the active elements, the overall training overhead to estimate the entire channel can be dramatically reduced.
To minimize the estimation complexity, the proposed technique is based on the linear combination of partial CSI, which only requires linear matrix operations.
By exploiting the spatial correlation among the RIS elements, proper weights for the linear combination and normalization factors are developed.
Numerical results show that the proposed technique outperforms other schemes using the active elements at the RIS in terms of the normalized mean squared error when the number of active elements is small, which is necessary to maintain the low cost and power consumption~of~RIS.
\end{abstract}


\section{Introduction}\label{sec1}
Reconfigurable intelligent surface (RIS) and its variations have been considered as a promising solution to achieve high energy efficiency and tackle the hardware cost issue for future wireless communication systems\cite{Zhang:2019a, Renzo:2019, Huang:2018}.
The RIS is a software-controllable meta-surface consisting of a large number of low-cost and passive reflecting elements integrated on a planar surface.
The RIS can control the amplitude and/or phase of the incoming signal in real-time, so a favorable channel response can be obtained \cite{Zhang:2019a, Renzo:2019}.

Recently, many researches on active and passive beamforming designs have been conducted to improve the performance of RIS-aided communication systems \cite{Zhang:2019b, Guo:2020}. However, to fully exploit the advantages of RIS-aided systems, the base station (BS) or user equipment (UE) needs to know accurate channel state information (CSI).
In general, the RIS is composed of purely passive elements and has no ability of receiving the training signals, which makes the BS or UE only observes the cascaded UE-RIS-BS channels.
However, the training overhead for the estimation of RIS-related channel increases with the number of RIS elements\cite{ICL:TWC}. When a large number of RIS elements are deployed to improve the performance of communication systems, the training overhead becomes the major bottleneck to deploy the RIS in practice.

One practical approach to reduce the training overhead for the channel estimation is to integrate the RIS with a small number of active elements, 
which have the capability of receiving and processing the training signals independently through receive radio frequency (RF) chains\cite{Taha:2021, Liu:2020, Jin:2021, Zhang:2021}.
When some RIS elements are replaced with the active elements, only partial CSI of the entire channel can be obtained through the active elements, which makes it necessary to develop new channel estimation schemes that acquire the entire channel from the partial CSI.
In \cite{Taha:2021},\cite{Liu:2020}, compressed sensing (CS) and deep learning-based channel estimation schemes were developed.
Moreover, the channel estimation in \cite{Jin:2021} utilized deep residual network with uniformly distributed active elements.
In \cite{Zhang:2021}, estimation signal parameter via rotational invariance technique (ESPRIT) was exploited to estimate channel parameters based on the correlation matrix obtained from received pilot signals at the active elements.
However, these channel estimation schemes have targeted millimeter-wave (mmWave) communication systems where only a small number of paths are dominant.
In sub-6 GHz channels, which will still be the major spectra for future wireless communications, new approaches considering a large number of paths need to be developed for the RIS-aided systems.

In this paper, we develop a novel channel estimation technique for the RIS-aided multi-user multiple-input single-output (MU-MISO) systems.
Considering a small number of active elements at the RIS, the proposed technique exploits the partial CSI obtained from the active elements. Specifically, the proposed technique linearly combines the partial CSI to estimate the entire channel.
By exploiting the spatial correlation among the RIS elements, proper weights for the linear combination are derived. In addition, normalization factors are introduced to make the norm of linearly combined estimate close to the norm of true channel.
The proposed technique only requires the pilot training for the partial CSI, making the overall training overhead irrelevant to the total number of passive elements at the RIS.
Numerical results show that the proposed technique outperforms the baseline schemes when the number of active elements is small, which is required for the low cost and power consumption of RIS.

The rest of this paper is organized as follows. 
Section \ref{sec2} presents the system model of RIS-aided MU-MISO system.
In Section~\ref{sec3}, we first explain the process of estimating the partial CSI through the pilot training. Then, the rank of partial CSI is analyzed, and the linear combination based channel estimation technique is proposed.
Numerical results for the channel estimation are provided in Section \ref{sec4}, and we conclude the paper in Section \ref{conclusion}.

\section{System Model}\label{sec2}
\begin{figure}
	\centering
	\includegraphics[width=1.02\columnwidth]{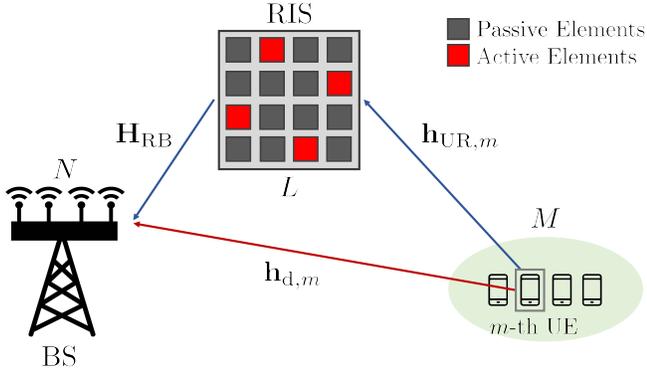}
	\caption{An example of the RIS-aided MU-MISO communication systems with $N$ BS antennas, $M$ single antenna UEs, and $L$ RIS elements.}
	\label{system_model}
\end{figure}
We consider a time division duplexing (TDD) RIS-aided MU-MISO communication system as shown in Fig. \ref{system_model}. 
The BS deploys $N$ antennas and serves $M$ single antenna UEs.
The RIS is assumed to have a uniform planar array (UPA) structure with $L_{\mathrm{h}}$ horizontal and $L_{\mathrm{v}}$ vertical elements where $L=L_{\mathrm{h}}L_{\mathrm{v}}$ denotes the total number of RIS elements.
The area of each RIS element is given by $A=d_{\mathrm{h}} d_{\mathrm{v}}$ with the horizontal width $d_{\mathrm{h}}$ and the vertical height $d_{\mathrm{v}}$.
Among all $L$ RIS elements, $L_{\mathrm{act}}$ active elements can operate in both sensing and reflecting modes with each RF chain. By exploiting the active elements, the RIS can receive and process the training signals, which can significantly reduce the training overhead\cite{IRS_tutorial}. The RIS is connected to the BS via a controller where the BS can control the RIS elements for desired signal reflection.

By focusing on uplink transmissions, the received signal at the BS is given by
\begin{equation}
	\by = \sum_{m=1}^M (\bh_{\mathrm{d},m}+ \bH_{\mathrm{RB}}\bTheta \bh_{\mathrm{UR},m})s_m + 
	\bn_{\mathrm{BS}},	 
\end{equation}
where $s_m$ is the transmit signal from the $m$-th UE satisfying $\mathbb{E}[\vert s_m \vert^2]  \leq P_m$ with the uplink transmit power $P_m$, and $\bn_{\mathrm{BS}} \sim \cC\cN(\b0, \sigma_{\mathrm{BS}}^2 \bI_N)$ is the additive white Gaussian noise (AWGN) vector at the BS with the noise variance $\sigma_{\mathrm{BS}}^2$.
The direct link channel from the $m$-th UE to the BS is denoted by $\bh_{\mathrm{d},m}\in \mathbb{C}^{N\times 1}$.
The uplink channel of the RIS-BS link is denoted by $\bH_{\mathrm{RB}} \in \mathbb{C}^{N \times L}$, and the channel from the $m$-th UE to the RIS is represented by  $\bh_{\mathrm{UR},m} \in \mathbb{C}^{L \times 1}$. The $L \times L$ reflection-coefficient matrix of the RIS is defined as $\bTheta= \mathrm{diag}\left([e^{j\theta_1},\cdots,e^{j\theta_L}]^{\mathrm{T}}\right)$.

Since the direct channel estimation can be viewed as a conventional MU-MISO channel estimation problem, we focus on the problem of RIS-related channel estimation.
Considering a large number of paths and the spatial correlation among the RIS elements, we adopt the correlated Rayleigh fading channel model for both the RIS-BS link and the UE-RIS links\cite{Bjornson:2021}. Moreover, the RIS is assumed to be deployed in an isotropic environment, which means that angle of departure (AoD) or angle of arrival (AoA) for both links are uniformly distributed.
Note that the RIS will usually support a set of closely located users, which may experience the same RIS correlation. Still, the channel between each UE and the RIS can be assumed to be linearly independent because the UEs will be geometrically separated by more than a few wavelengths \cite{Adhikary:2013}.
Then, the channel between the $m$-th UE and the RIS $\bh_{\mathrm{UR},m}$~is
\begin{equation}\label{h_UR}
	\bh_{\mathrm{UR},m} \sim \cC\cN(\b0, A\mu_{\mathrm{UR}}\bR), \enspace m=1,\cdots,M, 
\end{equation}
where $A\mu_{\mathrm{UR}}\bR$ is the $L \times L$ covariance matrix with the large-scale channel coefficient of UE-RIS links $\mu_{\mathrm{UR}}$, which can be modeled by the average signal attenuation.
If the BS antennas are well separated, each row of the RIS-BS channel $\bH_{\mathrm{RB}}$ can be similarly defined as (\ref{h_UR}) with the covariance matrix $A\mu_{\mathrm{BR}}\bR$ where $\mu_{\mathrm{BR}}$ denotes the large-scale channel coefficient of RIS-BS link.
The correlation among the RIS elements is expressed as the normalized spatial correlation matrix $\bR$.
In the isotropic scattering environment, the $(a,b)$-th entry of $\bR$ is \footnote{Although we adopted the model in \cite{Bjornson:2021} to represent the spatial correlation among the RIS elements, the proposed technique also works for other spatial correlation models that properly describe the relationship among the RIS elements such as the Kronecker model in \cite{Kronecker_1, Kronecker_2, Kronecker_3}.} \cite{Bjornson:2021}
\begin{equation} \label{correlation matrix}
	[\bR]_{a,b} = \mathrm{sinc}\left( \frac{2\Vert \bu_a - \bu_b \Vert}{\lambda} \right), \enspace a,b=1,\cdots,L,
\end{equation}
where $\mathrm{sinc}(x)=\frac{\mathrm{sin}(\pi x)}{\pi x}$ is the sinc function, $\lambda$ is the wavelength, and $\bu_a = [0, \enspace i(a)d_{\mathrm{h}}, \enspace j(a)d_{\mathrm{v}}]^{\mathrm{T}} $ is the location vector of the $a$-th element of RIS with $i(a) = \mathrm{mod}(a-1, L_{\mathrm{h}})$ and $j(a)=\lfloor (a-1)/L_{\mathrm{h}} \rfloor$. Note that mod($\cdot$,$\cdot$) represents the modulus operation, and  $\lfloor a \rfloor$ denotes the greatest integer less than or equal to the real number $a$.

\section{Proposed Channel Estimation Technique}\label{sec3}

In this section, we propose a channel estimation technique that estimates the entire channel from the partial CSI obtained from the active elements at the RIS.
The proposed technique exploits the correlation among the RIS elements and only requires linear operations, which makes the technique~practical.

Since the active elements in the RIS have capability of receiving and processing the training signals, the RIS-BS link channel and UE-RIS link channels can be estimated separately in a coherence time block.
Moreover, taking the advantage of channel reciprocity in the TDD system, the uplink RIS-BS channel can be obtained through the downlink BS-RIS channel \cite{Marzetta:2014}.
Hence, we only explain the estimation of UE-RIS link channels since the RIS-BS link channel estimation can be carried out similarly.
Unlike other existing schemes \cite{Zhang:2021, Jin:2021}, which estimate each UE-RIS channel separately, the proposed technique estimates the entire UE-RIS channels simultaneously.

\subsection{Estimation of UE-RIS sub-channels}\label{sec3_1}
For the pilot training, the active elements operate in the sensing mode to receive pilot signals from the UEs.
Each UE sends an orthonormal pilot sequence simultaneously during $\tau_{\mathrm{p}}$ time slots. Let $\bphi_m = [\phi_m(1), \cdots, \phi_m(\tau_{\mathrm{p}})]^{\mathrm{T}} \in \mathbb{C}^{\tau_{\mathrm{p}} \times 1}$ be the orthonormal pilot sequence sent by the $m$-th UE.
The signal received by the active elements at the $t$-th time slot is given~by
\begin{equation}
	\bx(t) = \sqrt{P_{\mathrm{UL}}}\sum_{m=1}^M \bar{\bh}_{\mathrm{UR},m} \phi_m(t) + \bn_{\mathrm{RIS}}(t),
\end{equation}
where $P_{\mathrm{UL}}$ is the uplink transmit power for pilot training, $\bar{\bh}_{\mathrm{UR},m} \in \mathbb{C}^{L_{\mathrm{act}}\times 1}$ is the sub-channel of $\bh_{\mathrm{UR},m}$ corresponding to the indices of active elements, and $\bn_{\mathrm{RIS}}(t) \sim \cC\cN(\b0, \sigma_{\mathrm{RIS}}^2 \bI_{L_\mathrm{act}})$ is the AWGN vector at the RIS with the noise variance $\sigma_{\mathrm{RIS}}^2$. By stacking the $\tau_{\mathrm{p}}$ received signals, the RIS obtains
\begin{align}
	\bX =[\bx(1),\cdots,\bx(\tau_{\mathrm{p}})]
	=\sqrt{P_{\mathrm{UL}}}\bH_{\mathrm{UR,act}}\bPhi^{\mathrm{T}} + \bN, 
\end{align}
where $\bH_{\mathrm{UR,act}}=\left[\bar{\bh}_{\mathrm{UR},1},\cdots,\bar{\bh}_{\mathrm{UR},M}\right]\in\mathbb{C}^{L_{\mathrm{act}}\times M}$ is the entire UE-RIS sub-channels, $\bPhi =[\bphi_1,\cdots,\bphi_M]\in~\mathbb{C}^{\tau_{\mathrm{p}} \times M}$ is the whole uplink pilot matrix, and $\bN=[\bn_{\mathrm{RIS}}(1),$ $\cdots, \bn_{\mathrm{RIS}}(\tau_{\mathrm{p}})] \in \mathbb{C}^{L_{\mathrm{act}}\times \tau_{\mathrm{p}}}$. We set $\tau_{\mathrm{p}} = M$ to take the minimum sequence length such that $\bPhi^{\mathrm{T}} \bPhi^* = \bI_M$. Then, the estimated UE-RIS sub-channels $\tilde{\bH}_{\mathrm{UR,act}}$ can be computed~as
\begin{align} \label{pilot_training}
	\tilde{\bH}_{\mathrm{UR,act}} = \frac{1}{\sqrt{P_{\mathrm{UL}}}}\bX \bPhi^* = \bH_{\mathrm{UR,act}} + \frac{1}{\sqrt{P_{\mathrm{UL}}}}\bN\bPhi^*.
\end{align}

\subsection{Rank of UE-RIS sub-channels}\label{sec3_2}
In this subsection, we discuss the rank of sub-channel matrix $\bH_{\mathrm{UR,act}}$ to develop the proposed technique that exploits the full rank property of $\bH_{\mathrm{UR,act}}$.
Denote the entire UE-RIS channels as $\bH_{\mathrm{UR}}=[\bh_{\mathrm{UR},1},\cdots,\bh_{\mathrm{UR},M}]$ and the covariance matrix of each UE-RIS channel as $\bK=A\mu_{\mathrm{UR}}\bR$.
By using the coloring transformation \cite{coloring_transform}, $\bH_{\mathrm{UR}}$ can be expressed as
\begin{equation}
	\bH_{\mathrm{UR}} = \bK^{\frac{1}{2}}\bZ,
\end{equation}
where each column of $\bZ \in \mathbb{C}^{L \times M}$ is independent and identically distributed (i.i.d.) with $\cC\cN(\b0, \bI_L)$. The entire UE-RIS sub-channels can be similarly written as
\begin{equation} \label{H_UR_act}
	\bH_{\mathrm{UR,act}} = \bK_{\mathrm{act}}^{\frac{1}{2}}\bZ,
\end{equation}
where $\bK_{\mathrm{act}}^{\frac{1}{2}} \in \mathbb{C}^{L_{\mathrm{act}} \times L}$ consists of $L_{\mathrm{act}}$ rows of $\bK^{\frac{1}{2}}$ corresponding to the indices of active elements. To analyze the rank of $\bH_{\mathrm{UR,act}}$, it is necessary to analyze the rank of $\bK_{\mathrm{act}}^{\frac{1}{2}}$.

The correlation coefficients in (3) imply that the difference of location vectors between two RIS elements determines how correlated they are, and the UPA structure of RIS indicates that the off-diagonal entries of $\bK$ having non-zero values always exist since only non-zero integer arguments make the sinc function become zero \cite{Bjornson:2021}. Along with the fact that $\bK$ is the symmetric matrix, it can be verified that $\bK$ is not the full rank matrix, and the positive semi-definiteness of $\bK$ indicates that the rank of $\bK^{\frac{1}{2}}$ is the same as the rank of $\bK$ \cite{matrix_analysis}.
Since some rows in $\bK^{\frac{1}{2}}$ may not be linearly independent, the rank of $\bK_{\mathrm{act}}^{\frac{1}{2}}$ depends on how many and which rows are selected, and these are determined by the number and the location of active elements.

\begin{figure}
	\centering
	\includegraphics[width=1.03\columnwidth]{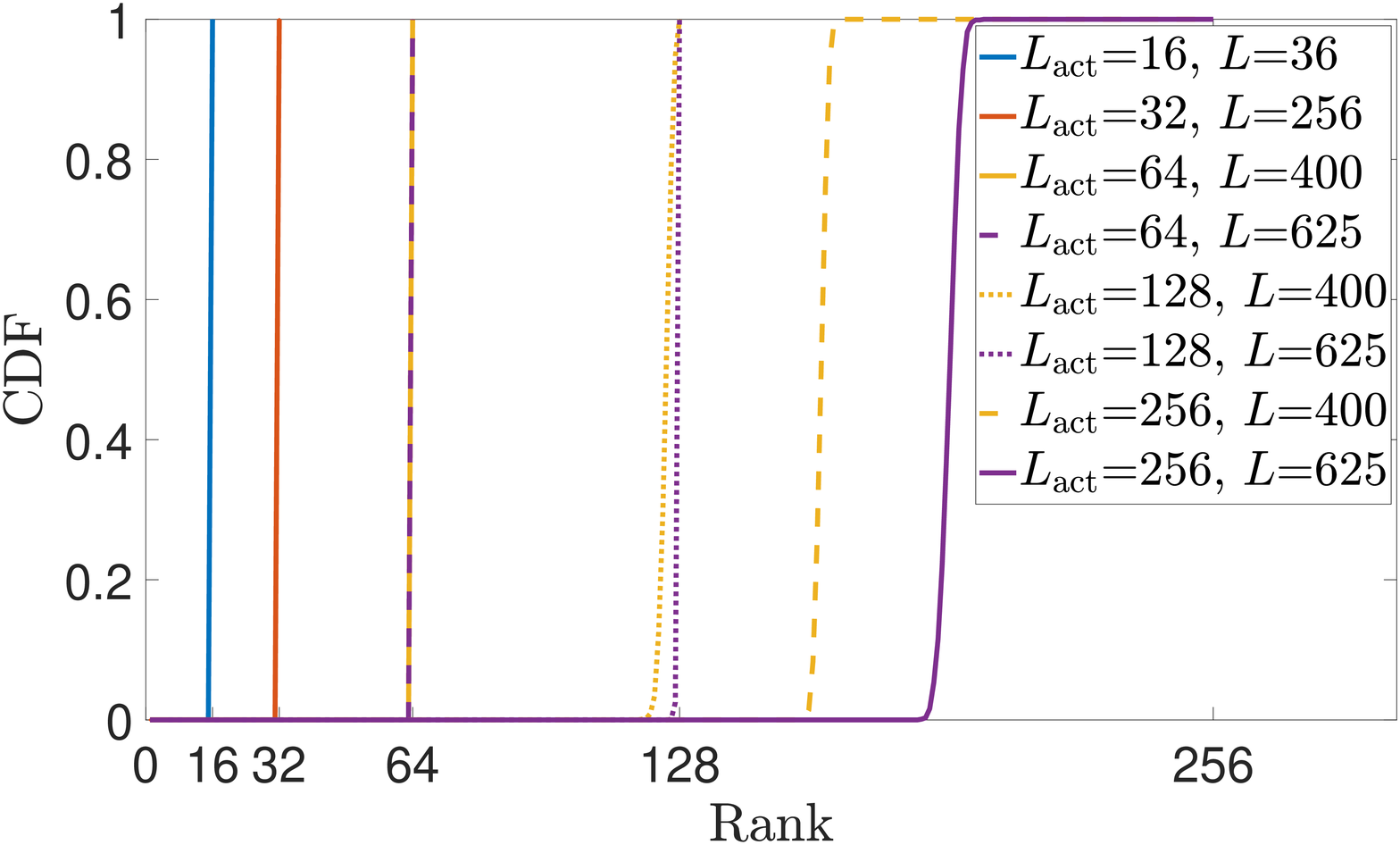}
	\caption{CDF of the rank of $\bK_{\mathrm{act}}^{\frac{1}{2}}$.}
	\label{fig:CDF}
\end{figure}
In Fig. \ref{fig:CDF}, cumulative distribution function (CDF) of the rank of $\bK_{\mathrm{act}}^{\frac{1}{2}}$ is plotted according to $L_{\mathrm{act}}$ and $L$ with $A\mu_{\mathrm{UR}}=$~$1$ and
randomly chosen locations of active elements at the RIS.
The first plot corresponds to the case where both $L_{\mathrm{act}}$ and $L$ are small. The next three plots correspond to the case where $L_{\mathrm{act}}$ is small, and $L$ is large. Full rank is guaranteed with high probability in these two situations. However, in the rest of plots, where both $L_{\mathrm{act}}$ and $L$ are large, full rank is not guaranteed. These numerical results imply that $\bK_{\mathrm{act}}^{\frac{1}{2}}$ is the full rank matrix with high probability when $L_{\mathrm{act}}$ is small compared to $L$. 
In general, the number of active elements $L_{\mathrm{act}}$ should be small to minimize the additional power consumption at the RIS. Hence, it is reasonable to assume that $\bK_{\mathrm{act}}^{\frac{1}{2}}$ is the full rank matrix.

Under the full rank assumption of $\bK_{\mathrm{act}}^{\frac{1}{2}}$, we further derive the rank of $\bH_{\mathrm{UR,act}}$ in the following proposition.
\begin{proposition}
	If $\bK_{\mathrm{act}}^{\frac{1}{2}}$ is the full rank matrix, $\bH_{\mathrm{UR,act}}$ is the full rank matrix.
	\begin{proof}
		Assume an $L \times L$ unitary matrix $\bU=[\bU_1 \enspace \bU_2]$ satisfying $\bU \bU^{\mathrm{H}}=\bI_L$, where the columns of $\bU_1 \in \mathbb{C}^{L \times L_{\mathrm{act}}}$ and $\bU_2 \in \mathbb{C}^{L \times (L- L_{\mathrm{act}})}$ correspond to the conjugate of orthonormal basis of $\mathrm{R}\left(\left(\bK_{\mathrm{act}}^{\frac{1}{2}}\right)^{\mathrm{T}}\right)$ and $\mathrm{N}\left(\bK_{\mathrm{act}}^{\frac{1}{2}}\right)$ with $\mathrm{R}(\cdot)$ and $\mathrm{N}(\cdot)$ representing the column and null spaces of a given matrix. Then, $\bH_{\mathrm{UR,act}}$ can be expressed as
		\begin{align} \label{H_UR_act_2}
			\bH_{\mathrm{UR,act}} 
			&= \bK_{\mathrm{act}}^{\frac{1}{2}}\bU\bU^{\mathrm{H}}\bZ =\bK_{\mathrm{act}}^{\frac{1}{2}}[\bU_1 \enspace \bU_2]\tilde{\bZ} \nonumber \\
			&=\left[\bK_{\mathrm{act}}^{\frac{1}{2}}\bU_1 \enspace \b0 \right]\tilde{\bZ}
			=\bK_{\mathrm{act}}^{\frac{1}{2}}\bU_1\hat{\bZ},
		\end{align}
		where $\tilde{\bZ}=\bU^{\mathrm{H}}\bZ$, which has the same distribution with $\bZ$, and $\hat{\bZ} \in \mathbb{C}^{L_{\mathrm{act}}\times M}$ consists of the first $L_{\mathrm{act}}$ rows of $\tilde{\bZ}$. Note that each column of $\hat{\bZ}$ is i.i.d. with $\cC \cN(\b0, \bI_{L_{\mathrm{act}}})$. If the first column of $\bU_1$ is set to the Hermitian of first row of $\bK_{\mathrm{act}}^{\frac{1}{2}}$ and the other columns are set to the orthonormal basis satisfying $\bU_1 \bU_1^{\mathrm{H}}=$~$\bI_{L}$, $\bK_{\mathrm{act}}^{\frac{1}{2}}\bU_1$ becomes the $L_{\mathrm{act}} \times L_{\mathrm{act}}$ lower triangular matrix.
		Since multiplication by a full-rank square matrix preserves the rank of a given matrix, the rank of (\ref{H_UR_act_2}) is equal to the rank of $\hat{\bZ}$, which is $\mathrm{min}(L_{\mathrm{act}}, M)$. Hence, $\bH_{\mathrm{UR,act}}$ is the full rank matrix when $\bK_{\mathrm{act}}^{\frac{1}{2}}$ is the full rank matrix, which finishes the~proof.
	\end{proof}
\end{proposition}

\subsection{Estimation of entire UE-RIS channels}\label{sec3_3}
In this subsection, we explain the proposed linear combination based channel estimation technique that reconstructs the entire channels $\bH_{\mathrm{UR}}$ from $\tilde{\bH}_{\mathrm{UR,act}}$.
From the rank analysis in Section \ref{sec3_2}, it is reasonable to assume that $\bH_{\mathrm{UR,act}}$ is the full rank matrix when $L_{\mathrm{act}}$ is small enough.
In the proposed technique, we assume $L_{\mathrm{act}} \geq M$, which can be easily satisfied in practice if the BS serves no more than $L_{\mathrm{act}}$ UEs at each coherence time block.	
Then, we can assume $\bH_{\mathrm{UR,act}}$ is the full column rank matrix, i.e., $\mathrm{rank}(\bH_{\mathrm{UR,act}})=M$. This implies that $M$ rows of $\bH_{\mathrm{UR,act}}$ form the row basis of $\bH_{\mathrm{UR}}$, and the other rows of $\bH_{\mathrm{UR}}$ can be expressed as the linear combination of $M$ rows in $\bH_{\mathrm{UR,act}}$, which is the foundation of our proposed technique.

The proposed channel estimation technique consists of two parts: 1) weighted linear combination and 2) normalization. We first explain the weighted linear combination part and the normalization part after.

\subsubsection{Weighted linear combination by exploiting correlation among rows}
Let $\cI_{\mathrm{act}}$ be the index set corresponding to the active elements of RIS.
To estimate the entire channels, the remaining $L-L_{\mathrm{act}}$ rows that do not correspond to the indices in $\cI_{\mathrm{act}}$ need to be estimated.
Assuming $\bH_{\mathrm{UR,act}}$ is the full rank matrix, based on the discussion in Section \ref{sec3_2}, regardless of the disposition of active elements, a reduced dimensional $M \times M$ matrix that consists of selected $M$ rows from $\bH_{\mathrm{UR,act}}$ is always the full rank matrix. This implies that arbitrary $M$ rows selected from $L_{\mathrm{act}}$ rows in $\bH_{\mathrm{UR,act}}$ become the row basis of $\bH_{\mathrm{UR}}$.

Let $\bH_{\mathrm{UR}}(\ell,:)$ be the $\ell$-th row of $\bH_{\mathrm{UR}}$. Although any $M$ rows can be used to estimate $\bH_{\mathrm{UR}}(\ell,:)$, it is desirable to select highly correlated rows with $\bH_{\mathrm{UR}}(\ell,:$$)$.
Since the correlation between the $a$-th entry and the $b$-th entry of $\bh_{\mathrm{UR},m}$ is represented by $[\bR]_{a,b}$ for $\forall m=1,\cdots,M$, $[\bR]_{a,b}$ is a good measure of the correlation between $\bH_{\mathrm{UR}}(a,:)$ and $\bH_{\mathrm{UR}}(b,:)$. Although $\bH_{\mathrm{UR}}(\ell,:)$ is not given but to be estimated, the correlation between $\bH_{\mathrm{UR}}(\ell,:)$ and the rows in $\bH_{\mathrm{UR,act}}$ can be obtained from $\bR$. Therefore, the proposed technique selects $M$ rows from $\tilde{\bH}_{\mathrm{UR,act}}$ having the largest correlation with $\bH_{\mathrm{UR}}(\ell,:)$ using $\bR$, and the corresponding correlation coefficients are utilized as the weights for the linear~combination. 

Although exact linear combination coefficients are hard to find for all rows to be estimated, we set the linear combination coefficients based on the correlation coefficients since highly correlated rows are likely to have relevant values according to their correlation. Among possible ways, we adopt the exponential weights to the correlation coefficients to give more weights to more correlated rows.
Let us define $\pi_1^{(\ell)},\cdots,\pi_{M}^{(\ell)}$ as the row indices corresponding to $M$ rows in $\tilde{\bH}_{\mathrm{UR,act}}$ having the largest correlation with $\bH_{\mathrm{UR}}(\ell,:)$ and $\psi_1^{(\ell)},\cdots,\psi_{M}^{(\ell)}$ as the row indices in $\bH_{\mathrm{UR}}$.
Then, the corresponding exponential weights are given~by
\begin{equation} \label{exponential_weight}
	w\left([\bR]_{\ell,\psi_m^{(\ell)}} \right) = \mathrm{sign}\left([\bR]_{\ell,\psi_m^{(\ell)}} \right)\mathrm{exp}\left( \alpha \left\vert[\bR]_{\ell,\psi_m^{(\ell)}}\right\vert \right),
\end{equation}
where the design parameter $\alpha$ denotes the weight coefficient, which would be numerically optimized.
For $\forall \ell \notin \cI_{\mathrm{act}}$, the estimate of $\bH_{\mathrm{UR}}(\ell,:)$ is then given by  
\begin{equation}
	\widehat{\bH}_{\mathrm{UR}}(\ell,:) = \sum_{m=1}^M w\left([\bR]_{\ell,\psi_m^{(\ell)}} \right)\tilde{\bH}_{\mathrm{UR,act}}\left(\pi_m^{(\ell)},: \right).
	\label{linear_combination}
\end{equation}

\subsubsection{Normalization of norm for estimated rows}
After applying (\ref{linear_combination}), it is necessary to make the norm of estimated channel close to that of the actual channel because the weighted linear combination of $M$ rows can cause significant difference for the norm value.
Based on a statistical distribution of norm of rows, we explain the normalization method.

Let us define a matrix $\bS=\sum_{m=1}^M \bh_{\mathrm{UR},m} \bh_{\mathrm{UR},m}^{\mathrm{H}}$.
Since all columns of $\bH_{\mathrm{UR}}$ are i.i.d. with $\cC\cN(\b0,\bK)$ where $\bK=$~$A\mu_{\mathrm{UR}}\bR$, $\bS$ follows the complex Wishart distribution, i.e., $\bS \sim \cC\cW_{L}(M, \bK)$, with $M$ degrees of freedom and the covariance matrix $\bK$ \cite{Morales:2011}.
The diagonal components of $\bS$ represent the squared norm values of rows of $\bH_{\mathrm{UR}}$, and the correlation coefficient between $\Vert\bH_{\mathrm{UR}}(a,:)\Vert^2$ and $\Vert\bH_{\mathrm{UR}}(b,:)\Vert^2$ is\cite{Ermolova:2012}
\begin{align}\label{correlation_norm}
	\rho_{a,b} = ([\bR]_{a,b})^2, \enspace \forall a,b=1,\cdots,L.
\end{align}
In (\ref{correlation_norm}), it is shown that the squared norms between two highly correlated rows are also highly correlated.
Since the RIS only knows $\tilde{\bH}_{\mathrm{UR,act}}$, to minimize the effect of noise in (\ref{pilot_training}), we adopt the normalization factor as the sample mean of the norm of rows used to perform the linear combination.
The normalization factor of the $\ell$-th estimated row $N_{\ell}$ is then given~by
\begin{equation}
	N_{\ell} = \frac{1}{M} \sum_{m=1}^M \left\Vert \tilde{\bH}_{\mathrm{UR,act}}\left(\pi_m^{(\ell)},: \right) \right\Vert.
\end{equation}
Finally, for $\forall \ell \notin \cI_{\mathrm{act}}$, the estimate of $\bH_{\mathrm{UR}}(\ell,:)$ applying the weighted linear combination and the normalization is given by
\begin{equation}
	\widehat{\bH}_{\mathrm{UR}}(\ell,:) = \frac{N_{\ell}\sum_{m=1}^Mw\left([\bR]_{\ell,\psi_m^{(\ell)}} \right)\tilde{\bH}_{\mathrm{UR,act}}\left(\pi_m^{(\ell)},: \right)}{\left\Vert \sum_{m=1}^Mw\left([\bR]_{\ell,\psi_m^{(\ell)}} \right)\tilde{\bH}_{\mathrm{UR,act}}\left(\pi_m^{(\ell)},: \right) \right\Vert}.
\end{equation}

\textit{Remark:} Throughout this section, we assumed a small number of $L_{\mathrm{act}}$ and relied on the full rank property of $\bH_{\mathrm{UR,act}}$. The proposed technique, however, still works with large $L_{\mathrm{act}}$, i.e., even when $\bH_{\mathrm{UR,act}}$ is not the full rank matrix, while there would be inevitable performance loss when estimating the channels corresponding to the passive elements. When $L_{\mathrm{act}}$ is large, it is possible to select highly correlated and independent $M$ rows to construct the full rank matrix and mitigate the loss with additional complexity.

\section{Numerical Results}\label{sec4}

In this section, we investigate the UE-RIS channel estimation performance of proposed linear combination based channel estimation technique. 
There are $M=8$ UEs in the same cluster.
The RIS deploys $L=16 \times 16$ elements with $d_{\mathrm{h}}=d_{\mathrm{v}}=\lambda/8$.
The UE-RIS distance is $d_{\mathrm{UR}}$ = 20 m.
With noise spectral density -174 dBm/Hz and bandwidth 1 MHz, the noise variance is set as $\sigma_{\mathrm{RIS}}^2=\mbox{-114 dBm}$.
The carrier frequency is set to be 3.5 GHz.
The pilot training sequence length for the UE-RIS sub-channel estimation is set as $\tau_{\mathrm{p}} = M$. The weight coefficient in (\ref{exponential_weight}) is numerically optimized as $\alpha=5$.

We compare the performance of proposed technique with the following baseline schemes:

\begin{itemize}
	\item Random coefficient : To estimate each row, randomly selected $M$ rows from $\tilde{\bH}_{\mathrm{UR,act}}$ are linearly combined with independently generated coefficients from $\cC\cN(0,1)$. 
	\item CS-based scheme \cite{Taha:2021} : Orthogonal matching pursuit (OMP) algorithm is used with the sparsity level $p$. The number of grid points in the azimuth and elevation dictionaries are set as $N_{\mathrm{D}}^{\mathrm{Az}}=2L_{\mathrm{h}}$ and $N_{\mathrm{D}}^{\mathrm{El}}=2L_{\mathrm{v}}$.
	\item ESPRIT based scheme \cite{Zhang:2021} : Total least square (TLS) ESPRIT and multiple signal classification (MUSIC) method are used.	
	This scheme is known to work well for sparse channels, e.g., mmWave channels.
\end{itemize}
Except for the ESPRIT-based scheme, the locations of active elements at the RIS are randomly chosen. 

\begin{figure}
	\centering
	\includegraphics[width=1.03\columnwidth]{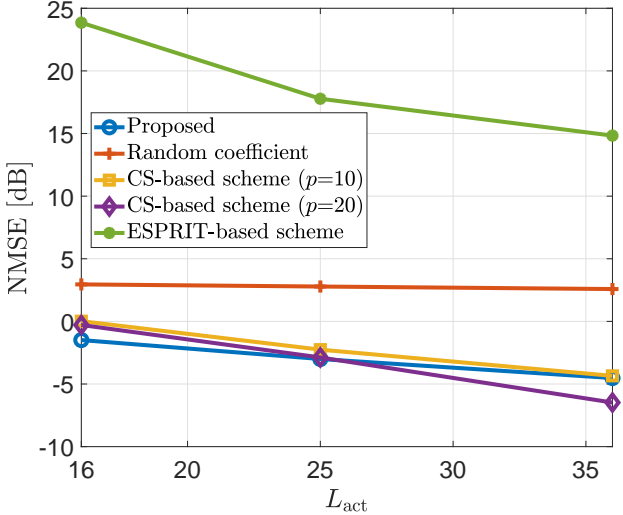}
	\caption{NMSE performance comparison according to the number of active elements.}
	\label{compare_active}
\end{figure}
Fig. \ref{compare_active} shows the normalized mean squared error (NMSE) according to the number of active elements $L_{\mathrm{act}}$ with the uplink pilot training power $P_{\mathrm{UL}}$ = 10 dBm, where the NMSE is defined as
\begin{equation}
	\mathrm{NMSE} =
	\frac{1}{L} \sum_{\ell=1}^L  \frac{ \Vert \bH_{\mathrm{UR}}(\ell,:) - \widehat{\bH}_{\mathrm{UR}}(\ell,:) \Vert^2}{\Vert \bH_{\mathrm{UR}}(\ell,:) \Vert^2}. \label{NMSE}
\end{equation} 
As $L_{\mathrm{act}}$ increases, the NMSE of proposed technique decreases due to the extended possibility of exploiting $M$ rows, which have much higher correlation among $L_{\mathrm{act}}$ rows in $\tilde{\bH}_{\mathrm{UR,act}}$ for the linear combination.
There is almost no performance gain in using random coefficients, which implies that utilizing highly correlated rows is crucial for the estimation performance.
When $L_{\mathrm{act}}$ is small, the proposed technique shows the lowest NMSE. As $L_{\mathrm{act}}$ increases, the CS-based scheme with $p=20$ shows lower NMSE than the proposed technique since large $p$ implies that the CS-based scheme considers more multipath components, and the estimated channel describes the true channel well.
However, the computational complexity of CS-based scheme is $\cO(ML_{\mathrm{act}}N_{\mathrm{D}}^{\mathrm{Az}}N_{\mathrm{D}}^{\mathrm{El}}p)$, and it linearly increases with $p$ and the dictionary size.
The computational complexity of our proposed technique is $\cO(2(L-L_{\mathrm{act}})M^2)$, and it would be significantly lower than the CS-based scheme in general since a larger dictionary size is required for the CS-based scheme as the number of RIS elements increases to guarantee the estimation performance.
It is observed that the ESPRIT-based scheme shows the worst NMSE performance since the maximum number of paths that can be estimated is limited according to the size of sub-surface, and the ESPRIT-based scheme cannot operate properly due to the small number of active elements in the sub-6 GHz spectra.

\begin{figure}
	\centering
	\includegraphics[width=1.03\columnwidth]{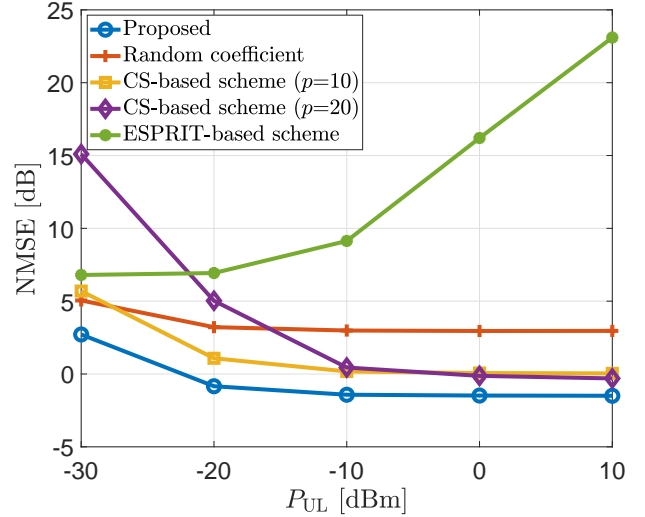}
	\caption{NMSE performance comparison according to the uplink transmit power.}
	\label{compare_power}
\end{figure}
Fig. \ref{compare_power} compares the NMSE with respect to $P_{\mathrm{UL}}$ for fixed $L_{\mathrm{act}}=16$.
The proposed technique shows the lowest NMSE regardless of the uplink transmit power for pilot training, and except for the ESPRIT-based scheme, the overall trends are the same as those in Fig. \ref{compare_active}.
It can be observed that the NMSE curves have constant slopes. This is because there is inevitable error caused by the estimation of channels corresponding to the passive elements even with accurate $\bH_{\mathrm{UR,act}}$ at high $P_{\mathrm{UL}}$.
The NMSE of ESPRIT-based scheme increases with $P_{\mathrm{UL}}$ since the reduced effect of noise makes the impact of a large number of paths in the channel noticeable, which reveals more mismatch with the environment for the ESPRIT-based scheme to work~well.

\section{Conclusion}\label{conclusion}
In this paper, we proposed a novel channel estimation technique with short training sequence length and low complexity in RIS-aided multi-user systems.
The proposed technique exploits the full rank structure of sub-channel corresponding to the active elements when the number of active elements is sufficiently small compared to the total number of RIS elements. The proposed technique performs the linear combination to estimate the entire channels by exploiting the correlation matrix among the RIS elements, where exponential weights and normalization factors are developed.
Numerical results verified that the proposed technique outperforms the baseline schemes in terms of the NMSE when the number of active elements is small, which is necessary to maintain the low cost and power consumption of RIS.

\bibliographystyle{IEEEtran}
\bibliography{refs_all}

\end{document}